\documentclass[conference,a4paper]{IEEEtran}
\IEEEoverridecommandlockouts
\usepackage{cite}
\usepackage{bm}
\usepackage[cmex10]{amsmath} 
\usepackage{extarrows}
\usepackage{amssymb}
\usepackage{graphicx}
\usepackage{color}
\usepackage{enumerate}
\usepackage{bookmark} 
\graphicspath{{figure}}
\usepackage{setspace}
\usepackage{subfigure}
\usepackage{algorithm}
\usepackage{algorithmicx}
\usepackage{multirow}
\usepackage{stfloats}
\usepackage{graphbox}
\usepackage{bbm}
\usepackage[utf8]{inputenc} 
\usepackage[T1]{fontenc}
\usepackage{url}
\usepackage{booktabs} 
\usepackage{graphicx}
\usepackage{mleftright}       % fix to wrong spacing of \left-,
\mleftright                   % \middle- \right-commands 
\usepackage{framed}  
\usepackage{scalerel}

\newcommand{\mr}{\mathrm} 

\newcommand{\bb}{\mathbbm}
\newcommand{\BE}{\begin{equation}}
\newcommand{\EE}{\end{equation}}
\newcommand{\BS}{\begin{subequations}}
\newcommand{\ES}{\end{subequations}}
\renewcommand{\bf}{\bm}

\newtheorem{theorem}{Theorem}

\newtheorem{lemma}{Lemma}

\newcommand{\tabincell}[2]{\begin{tabular}{@{}#1@{}}#2\end{tabular}}
\ifCLASSINFOpdf
\graphicspath{{figure/}}

\else

\fi

\begin{document}

\title{{Generalized Linear Systems with OAMP/VAMP Receiver: Achievable Rate and Coding Principle}}
\author{%
  \IEEEauthorblockN{Lei Liu\IEEEauthorrefmark{1}, \emph{Member, IEEE}, 
                    Yuhao Chi\IEEEauthorrefmark{2}, \emph{Member, IEEE}, \\
                    Ying Li\IEEEauthorrefmark{2}, \emph{Member, IEEE},
                    and Zhaoyang Zhang\IEEEauthorrefmark{1}, \emph{Senior Member, IEEE}}
  \IEEEauthorblockA{\IEEEauthorrefmark{1}%
                    College of Information Science and Electronic Engineering, Zhejiang University, China}
  \IEEEauthorblockA{\IEEEauthorrefmark{2}%
                    State Key Lab of ISN, Xidian University, China}
}

\maketitle
\vspace{-1mm}
\begin{abstract}
The generalized linear system (GLS) has been widely used in wireless communications to evaluate the effect of nonlinear preprocessing on receiver performance. Generalized approximation message passing (AMP) is a state-of-the-art algorithm for the signal recovery of GLS, but it was limited to measurement matrices with independent and identically distributed (IID) elements. To relax this restriction, generalized orthogonal/vector AMP (GOAMP/GVAMP) for unitarily-invariant measurement matrices was established, which has been proven to be replica Bayes optimal in uncoded GLS. However, the information-theoretic limit of GOAMP/GVAMP is still an open challenge for arbitrary input distributions due to its complex state evolution (SE). To address this issue, in this paper, we provide the achievable rate analysis of GOAMP/GVAMP in GLS, establishing its information-theoretic limit (i.e., maximum achievable rate). Specifically, we transform the fully-unfolded state evolution (SE) of GOAMP/GVAMP into an equivalent single-input single-output variational SE (VSE).  Using the VSE and the mutual information and minimum mean-square error (I-MMSE) lemma, the achievable rate of GOAMP/GVAMP is derived. Moreover, the optimal coding principle for maximizing the achievable rate is proposed, based on which a kind of low-density parity-check (LDPC) code is designed. Numerical results verify the achievable rate advantages of GOAMP/GVAMP over the conventional maximum ratio combining (MRC) receiver based on the linearized model and the BER performance gains of the optimized LDPC codes ($0.8 \sim 2.8$~dB) compared to the existing methods.
\end{abstract}

\vspace{-1mm}
\section{Introduction}
In wireless communication applications, a generalized linear system (GLS) has been widely adopted to solve the recovery problem of an unknown signal $\bf{x}$ from a nonlinear noisy observation $\bf{y}$ in the following form:
\BE\label{Eqn:GLS}
\bf{y}=Q(\bf{A}\bf{x}, \bf{n}), 
\EE
where $\bf{A}$ is the measurement matrix, $\bf{n}$ is an additive white Gaussian noise (AWGN), and $Q(\cdot)$ is a nonlinear function. Compared to the most commonly used standard linear system (SLS), i.e., $\bf{y} = \bf{A}\bf{x} + \bf{n}$, GLS can better account for the effect of nonlinear preprocessing on signal recovery in practical transceivers~\cite{MIMOADC2009,clip_cl, Phase_rec_SPL, Phase_MIMO_twc, Phase_retrieval_tsp}. For example, a quantization is necessary for the receiver to reduce the hardware cost and power efficiency of massive multiple-input multiple-output (MIMO) systems\cite{MIMOADC2009}. To suppress the high peak-to-average ratio (PAPR) induced by multi-carrier signal transmission in orthogonal frequency-division multiplex (OFDM), signal pre-distortion such as peak-clipping operation $Q(\cdot)$ is used to decrease envelope fluctuations of OFDM signals~\cite{clip_cl}. In a nutshell, these applications require the recovery of complete transmission signals based on the received nonlinear noisy observations and a specific nonlinear preprocessing function.

\vspace{-0.2cm}
\subsection{Linearized Approximate Model}
Despite the widespread use of GLS, an exact analysis is difficult due to the nonlinear nature of $Q(\cdot)$. Several linearized approximation models have been developed over the last decade to convert the GLS into a simplified SLS. Quantization noise is assumed to be additive and independent in \cite{Sundeep2015ADC} based on the additive quantization noise model (AQNM), based on which the achievable rate is calculated for the MRC receiver in massive MIMO with Gaussian signaling and low-resolution ADCs\cite{wck2015ADC}.  Similarly,  a linearized clipping model is developed for clipped GLS, and an iterative soft compensation method is proposed to mitigate the clipping distortion \cite{ShansuoWCL2019}. Even though linearization can significantly simplify performance analysis and algorithm design for GLS, it is intrinsically suboptimal, resulting in considerable degradation of bit-error-rate (BER) or achievable rate performance.

\vspace{-0.2cm}
\subsection{Bayesian Algorithms}
In the past few years, various generalized approximate message passing (AMP)-type algorithms based on the Bayesian framework have been developed for GLS with continuing development of AMP-type algorithms in SLS. A generalized AMP (GAMP) is proposed for GLS in \cite{GAMP2011}, which is low-complexity but is limited to IID matrices. To overcome the limitation of GAMP, generalized vector AMP (GVAMP) \cite{GVAMP2016} is proposed for GLS with unitarily-invariant  matrices. Meanwhile, GVAMP is proved to be replica Bayes optimal \cite{SundeepJSIT}. Due to the equivalence of GVAMP and generalized orthogonal AMP (GOAMP), they are referred to as GOAMP/GVAMP in this paper. By the utilization of a high complexity linear minimum mean squared error (LMMSE) to reduce linear interference, GOAMP/GVAMP has a high computational complexity. Recently, a low-complexity generalized memory AMP (GMAMP) is extended by MAMP \cite{LeiMAMP} for GLS with unitarily-invariant matrices \cite{GMAMP2022}, in which the Bayes optimality of GMAMP is also proven via SE. However, the results in \cite{GAMP2011, GVAMP2016, SundeepJSIT, GMAMP2022} are limited to the uncoded GLS, where error-free performance is not guaranteed. To the best of our knowledge, a rigorous investigation of the information-theoretic limit of generalized AMP-type algorithms in the GLS is yet lacking.

\subsection{Information-Theoretical Limits of AMP-type Algorithms}
The information-theoretical (i.e., constrained capacity) optimality of AMP for a coded SLS with an arbitrary input distribution and IID matrices is proven in \cite{LeiTIT2021}. That is, the achievable rate of AMP has been rigorously proven to be equal to the constrained capacity of SLS as derived in \cite{LScapacity1, LScapacity2}. The optimal coding principle is also provided for AMP based on the matching principle between the transfer functions of the linear detector (LD) and nonlinear detector (NLD). The constrained-capacity optimality of OAMP/VAMP for a coded SLS is proven for right unitarily-invariant matrices \cite{LeiOptOAMP} and is proven to achieve the constrained-capacity region of multi-user (MU) MIMO systems \cite{YuhaoTcom2022}. Meanwhile, the optimal coding principle is presented for OAMP/VAMP in~\cite{LeiOptOAMP, YuhaoTcom2022}.

However, the achievable rate analysis of the AMP-type algorithms\cite{LeiTIT2021, LeiOptOAMP, YuhaoTcom2022} in SLS cannot be directly applied to the generalized AMP-type algorithms in GLS. The reason is that the achievable rate analysis of the AMP-type algorithms relies on the single-input-single-output (SISO) transfer functions in their SEs. In contrast, the LD transfer function of the generalized AMP-type algorithms is dual-input-dual-output (DIDO) and is coupled with those of two NLDs, making the achievable rate analysis much more complicated. 

\subsection{Contributions of This Paper}
In this paper, we present the information-theoretical (i.e., achievable rate) analysis of  GOAMP/GVAMP in the coded GLS. To circumvent the difficulty of SE analysis of GOAMP/GVAMP, a multi-layer information matching principle is proposed to analyze the asymptotic performance of GOAMP/GVAMP. Specifically, a transfer function of the enhanced LD (ELD) is established for GOAMP/GVAMP. Then,  GOAMP/GVAMP can be asymptotically characterized by a variational SE (VSE) consisting of the transfer functions of ELD and NLD. Using the VSE, the achievable rate of GOAMP/GVAMP is derived based on the mutual information and MMSE (I-MMSE) lemma \cite{GuoTIT2005}. Moreover, we study the maximum achievable rate and practical LDPC code design of GOAMP/GVAMP in the coded GLS with clipping. The main contributions of this paper are summarized as follows.
\begin{enumerate}
    \item The variational SE is proposed to analyze the achievable rate of GOAMP/GVAMP in GLS, based on which the optimal coding principle is developed to maximize the achievable rate of GOAMP/GVAMP.
    \item The maximum achievable rate of GOAMP/GVAMP in a coded GLS is discussed for clipping. To validate its advantages, the GOAMP/GVAMP with optimized coding is compared to the conventional MRC receiver based on the linearized model.
    \item A kind of irregular LDPC code is designed for GOAMP/GVAMP with the aim of maximizing the achievable rate. Numerical results show that the finite-length performances of the optimized LDPC codes and quadrature phase-shift keying (QPSK) modulation are within $1.0$~dB from the threshold limits and outperform those of the existing state-of-art methods.
\end{enumerate}
% \emph{Note:} Due to the limitation of pages, the detailed discussions for achievable rate analysis of GOAMP/GVAMP with quantization and general high-dimensional neural network are given in a full version \cite{LeiGOAMP} of this paper.
\subsection{Notations}
For simplicity, we define a ``circle minus" operation $\bf{a} \ominus_c \bf{b} \equiv \tfrac{1}{1-c}(\bf{a}-c\bf{b})$ for estimate orthogonalization, where $c$ denotes the orthogonalization coefficient. Define a ``box minus" operation $a \boxminus b \equiv (a^{-1}-b^{-1})^{-1}$ for variance orthogonalization. Define $\langle\bf{A}_{M\times N}, \bf{B}_{M\times N}\rangle$ $ \equiv$ $\bf{A}_{M\times N}^{\rm H}\bf{B}_{M\times N}$ and  $\langle\bf{A}_{M\times N}| \bf{B}_{M\times N}\rangle \equiv$$\tfrac{1}{N}\langle\bf{A}_{M\times N}, \bf{B}_{M\times N}\rangle$. 
\vspace{-0.1cm}
\section{System Model}
\begin{figure}[t!]%\vspace{-0.4cm}
	\centering
	\includegraphics[width=0.9\columnwidth]{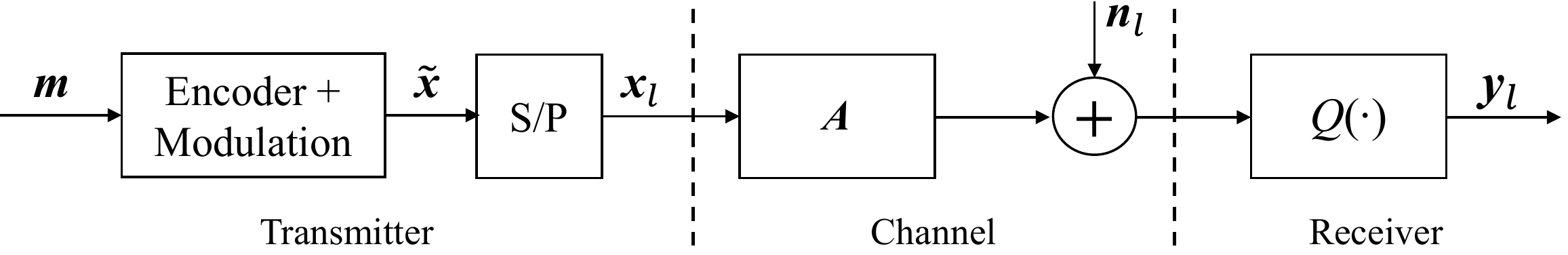}\vspace{-0.2cm}
	\caption{Illustration of a GLS, where S/P denotes serial-to-parallel conversion, $\bm{A}$  the channel matrix, and ${Q}(\cdot)$  the nonlinear preprocessing at the receiver.}\label{Fig:CGLM}  \vspace{-0.5cm}
\end{figure}

Fig.~\ref{Fig:CGLM} illustrates a coded GLS with an $M$-antenna transmitter and one receiver equipped with $N$ antennas. At the transmitter, a message vector $\bm{m}$ is encoded and modulated to a length-$NL$ symbol vector $\bm{\tilde{x}}$. Each element of $\bm{\tilde{x}}$ is taken independently from a constellation set $\mathcal{S}$ and $\bm{\tilde{x}}$ is split into $L$ length-$N$ vectors  $\{\bm{{x}}_l, l= 1, ..., L\}$ by serial-to-parallel conversion, which are transmitted into the linear channel through $L$ time slots. We assume that $\bm{x}_l$ satisfies the power constraint $\frac{1}{N}E\{||\bm{x}_l||^2\}=1$.

The receiver obtains signal $\bm{y}_l \in \mathbb{C}^{M\times 1}$ is expressed as  
\BE\label{Eqn:RecvSig}
\bm{y}_l ={Q}(\bm{A}\bf{x}_l +\bm{n}_l),
\EE
where $\bm{A}\in \mathbb{C}^{M\times N}$ is a quasi-static channel matrix, $\bm{n}_l\sim \mathcal{CN}(\mathbf{0},\sigma^2\bm{I})$  an additive white Gaussian noise (AWGN), and ${Q}(\cdot)$  a symbol-by-symbol nonlinear function. Without loss of generality, we assume $\tfrac{1}{\mathcal{J}}{\rm tr}\{\bf{A}^{\rm{H}}\bf{A}\}=1$, $\mathcal{J}={\mr{max}}\{M, N\}$, and the signal-to-noise ratio (SNR) is defined as  ${snr} = \sigma^{-2}$. The goal is to recover the message vector $\bm{m}$ based on $\bm{y}$, ${Q}(\cdot)$, $\bm{A}$ and the distribution of $\bm{x}_l$, 
\vspace{-0.1cm}
\section{Achievable Rate Analysis of GOAMP/GVAMP}
\subsection{GOAMP/GVAMP Receiver}
Since the GLS detection in each time slot is the same, we omit the time index $l$ in the rest of this paper for simplicity. For simplicity of discussion, we rewrite the GLS model as:
\BS\label{Eqn:model_CGLS}
\begin{align}
\Psi:\quad   &\bf{y}= {Q}(\bf{z}+\bf{n})\vspace{-2mm},\label{Eqn:CGLS_1}\\
\Gamma:\quad &\bf{z}=\bf{A}\bf{x}\vspace{-2mm},\label{Eqn:CGLS_2}\\
\Phi_{\mathcal{C}}: \quad &\bf{x} \in \bf{\mathcal{C}}\;\;{\rm and} \;\; x_i \sim P_X(x_i), \forall i.\label{Eqn:CGLS_3}
\end{align}
\ES
 
\begin{figure}[t] \vspace{-0.5cm}
\centering  
\subfigure[GOAMP/GVAMP receiver: $\hat{\psi}_t$ $\hat{\gamma}_t$  and $\hat{\phi}_t$ denotes the local MMSE/LMMSE detection for the local constraints $\Psi$, $\Gamma$ and $\Phi$, respectively. Orth denotes the orthogonal operations.]{
\includegraphics[width=0.8\columnwidth]{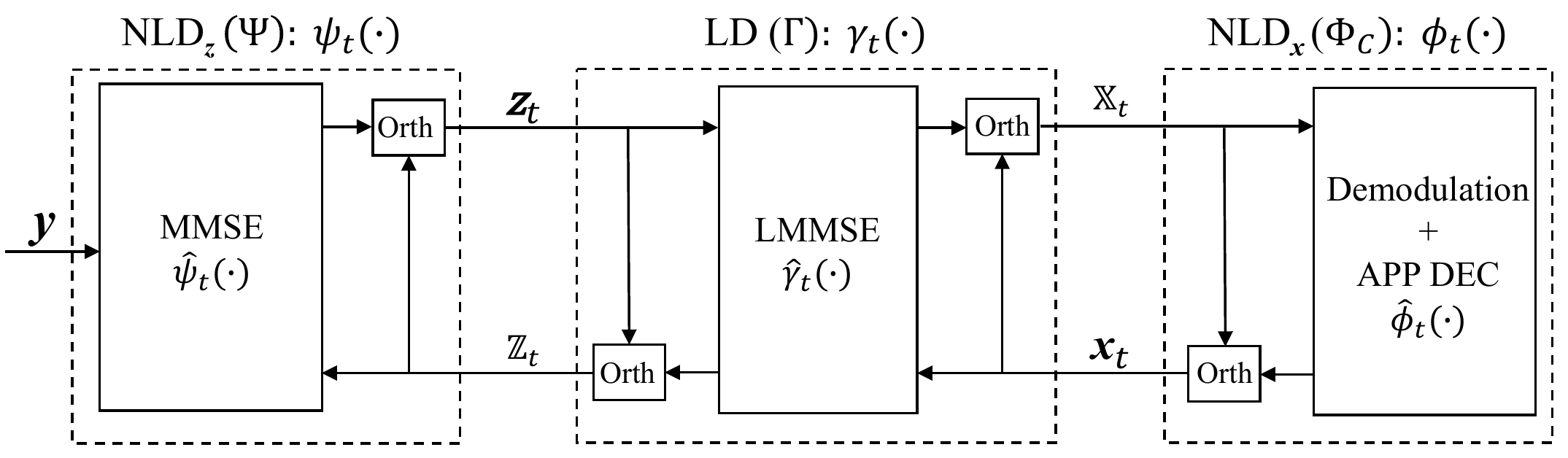}\label{Fig:GOAMP_iter}
}
\centering 
\subfigure[State evolution of GOAMP/GVAMP: $\psi_{\rm{SE}}$, $\phi_{\rm{SE}}$ and $\gamma_{\rm{SE}}$ denote MSE transfer functions of $\psi_t$, $\phi_t$ and $\gamma_t$, respectively.]{ 
\includegraphics[width=0.8\columnwidth]{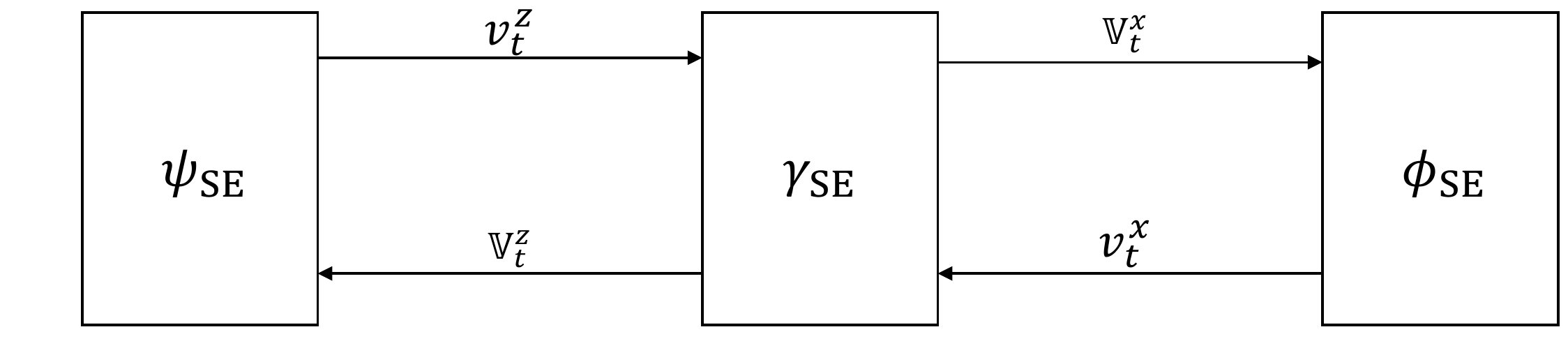}\label{Fig:GOAMP_SE}
} \vspace{-0.2cm}
\caption{Illustration of the GOAMP/GVAMP receiver and its state evolution.} \label{Fig:GOAMP}\vspace{-0.3cm}
\end{figure}

Fig.~\ref{Fig:GOAMP_iter} shows that the GOAMP/GVAMP receiver consists of an LD and two NLDs, where LD employs LMMSE detection for linear constraint $\Gamma$ in \eqref{Eqn:CGLS_2}, ${\text{NLD}}_z$ employs MMSE detection for nonlinear constraint $\Psi$ in \eqref{Eqn:CGLS_1}, and ${\text{NLD}}_x$ employs MMSE demodulation and \emph{a-posteriori probability} (APP) decoding for coding constraint $\Phi_{\mathcal{C}}$ in \eqref{Eqn:CGLS_3}.
%\vspace{-0.3cm}

\emph{GOAMP/GVAMP Receiver}: Starting with $t=1$ and $\bf{\bb{z}}_1=\bf{\bb{x}}_1=0$,
\BS\label{Eqn:GOAMP}
\begin{align}
\!\!\!\!\!\!\text{NLD:}&
\left[ \begin{aligned} {{{\bm{x}}_t}} \\ {{{\bm{z}}_t}} \end{aligned} \right] \!=\! 
\left[ \begin{aligned} {{\phi}_t(\bf{\bb{x}}_t)}\\ {\psi}_t(\bf{\bb{z}}_t) \end{aligned} \right] \!=\! \left[ \begin{aligned}  
\hat{\phi}_t(\bf{\bb{x}}_t) \ominus_{c^{\phi}_t} \bf{\bb{x}}_t\\
\hat{\psi}_t\big(\bf{\bb{z}}_t) \ominus_{c^{\psi}_t} \bf{\bb{z}}_t
 \end{aligned} \right], \label{Eqn:NLD}\\
\!\!\!\!\!\!\text{LD:}& \left[ \begin{aligned}  {\bf{\bb{x}}_{t+1}} \\ {\bf{\bb{z}}_{t+1}} \end{aligned} \right]\!=\!
\left[ \begin{aligned} {\gamma}^x_t({\bm{x}}_t, {\bm{z}}_t)\\{\gamma}^z_t({\bm{x}}_t, {\bm{z}}_t)  \end{aligned} \right] \!=\!\left[ \begin{aligned}
&\hat{\gamma}_t({\bm{x}}_t, {\bm{z}}_t)\ominus_{1-\delta c^{\gamma}_t} {\bm{x}}_t \\
&\bf{A}\hat{\gamma}_t({\bm{x}}_t, {\bm{z}}_t) \ominus_{c^{\gamma}_t} {\bm{z}}_t
\end{aligned}\right],
\label{Eqn:LD}
\end{align}
\ES
where ${{{\bm{x}}_t}}=[{{{x}}_{t,1}}, ..., {{{{x}}_{t,N}}}]^T$ and ${{{\bm{z}}_t}}=[{{{z}}_{t,1}}, ..., {{{{z}}_{t,M}}}]^T$ denote the outputs of ${{\phi}_t(\bf{\bb{x}}_t)}$ and ${{\psi}_t(\bf{\bb{z}}_t)}$ respectively, $\bf{\bb{x}}_t=[{\bf{\bb{x}}_{t,1}}, ..., {\bf{\bb{x}}_{t,N}}]^T$ and $\bf{\bb{z}}_t=[{\bf{\bb{z}}_{t, 1}}, ..., {\bf{\bb{z}}_{t, M}}]^T$ the outputs of ${\gamma}_t^x({\bm{x}}_t, {\bm{z}}_t)$ and ${\gamma}_t^z({\bm{x}}_t, {\bm{z}}_t)$, and superscripts $\psi$, $\phi$ and $\gamma$ correspond to the constraints $\Psi$, $\Phi_{\mathcal{C}}$ and  $\Gamma$, respectively.

\emph{NLD}: The local MMSE functions of ${{\psi}_t(\bf{\bb{z}}_t)}$ and ${{\phi}_t(\bf{\bb{x}}_t)}$ in~\eqref{Eqn:NLD} are given by
\BE \label{Eqn:local_nld}
 \hat{\phi}_t(\bf{\bb{x}}_t) \equiv \mr{E}\{ \bf{x} |\bf{\bb{x}}_t, \Phi\}, \quad 
 \hat{\psi}_t(\bf{\bb{z}}_t) \equiv \mr{E}\{ \bf{z} |\bf{\bb{z}}_t, \Psi\},
\EE
and %$c^{\phi}_t$ and $c^{\psi}_t$ are
%\BE\nonumber %\label{Eqn:local_nld_para}
%\begin{align}
 $c^{\phi}_t=\tfrac{1}{N{\bb{v}}_t^x}\mr{E}\{||\hat{\phi}_t(\bf{\bb{x}}_t)-\bm{x}||^2\}$ and 
 $c^{\psi}_t=\tfrac{1}{M{\bb{v}}_t^z}
 \mr{E}\{||\hat{\psi}_t(\bf{\bb{z}}_t) -\bm{z}||^2\}$, 
%\end{align}
%\EE
where ${\bb{v}}_t^x$ and ${\bb{v}}_t^z$ denote the input  variances of  ${{\phi}_t(\bf{\bb{x}}_t)}$ and ${{\psi}_t(\bf{\bb{z}}_t)}$ from ${\gamma}_t^x({\bm{x}}_t, {\bm{z}}_t)$ and ${\gamma}_t^z({\bm{x}}_t, {\bm{z}}_t)$, respectively. It is noted that the decoder $\hat{\phi}_t(\cdot)$ is assumed to be Lipschitz-continuous in this paper.

\emph{LD}: The local LMMSE function $\hat{\gamma}_t({\bm{x}}_t, {\bm{z}}_t)$ in~\eqref{Eqn:LD} is
% \BS
\BE \label{Eqn:local_ld}
\hat{\gamma}_t({\bm{x}}_t, {\bm{z}}_t) \equiv \bm{x}_t + \bf{A}^{\rm{H}}(\rho_t\bf{I}+\bf{A}\bf{A}^{\rm{H}})^{-1}(\bm{z}_t-\bf{A}\bm{x}_t),
\EE 
with $\rho_t={v_t^z}/{v_t^x}$ and  
% \BE\label{Eqn:local_ld_para}
$c^{\gamma}_t = \tfrac{1}{M}{\mr{Tr}}\{\bf{A}^{\rm{H}}(\rho_t\bf{I}+\bf{A}\bf{A}^{\rm{H}})^{-1}\bf{A}\}$,
% \EE\ES
where $v_t^x$ and $v_t^z$ denote the input variances of ${\gamma}_t^x({\bm{x}}_t, {\bm{z}}_t)$ and ${\gamma}_z^x({\bm{x}}_t, {\bm{z}}_t)$ from ${{\phi}_t(\bf{\bb{x}}_t)}$ and ${{\psi}_t(\bf{\bb{z}}_t)}$, respectively.

\emph{Note:} The $\hat{\gamma}_t(\cdot)$ and $\hat{\psi}_t(\cdot)$ have been proven to be Lipschitz-continuous in \cite{Takeuchi2020,GMAMP2022}. Meanwhile, the LDPC decoder $\hat{\phi}_t(\cdot)$ is proved to be Lipschitz-continuous in \cite[Appendix B]{ebert2023sparse}, indicating that the SE of GOAMP/GVAMP based on LDPC decoding holds. Although there is no strict proof for other types of FEC codes, we conjecture that $\hat{\phi}_t(\cdot)$ is also Lipschitz-continuous for the majority of FEC codes (e.g., Turbo code, Polar code, etc.).

\vspace{-0.3cm}
\subsection{State Evolution (SE)}
Based on the asymptotic IID Gaussianity lemma (see \cite{GMAMP2022} for more details), the asymptotic MSE of GOAMP/GVAMP can be characterized by the following SE. 
% \vspace{-1cm}
 \BS\label{Eqn:SE_GOAMP}
\begin{align}
\text{NLD:}\quad &
\left[ \begin{aligned}   v_t^x \\ v_t^z \end{aligned} \right] = 
\left[ \begin{aligned}  {{\phi}_{\mr{SE}}({\bb{v}}_t^x)}\\ {{\psi}_{\mr{SE}}({\bb{v}}_t^z)}  \end{aligned} \right] = \left[ \begin{aligned} 
\hat{v}_t^x\boxminus {\bb{v}}_t^x\\
 \hat{v}_t^z\boxminus {\bb{v}}_t^z
\end{aligned} \right], \label{Eqn:SENLD}\\
\text{LD:}\quad & \left[ \begin{aligned} {\bb{v}}_{t+1}^x \\ {\bb{v}}_{t+1}^z \end{aligned} \right] = \left[ \begin{aligned} {\gamma}_{\mr{SE}}^x(v_{t}^x, v_{t}^z) \\ {\gamma}_{\mr{SE}}^z(v_{t}^x, v_{t}^z) \end{aligned} \right] =
\left[ \begin{aligned}
& \hat{\bb{v}}_t^x \boxminus v_t^x\\
& \hat{\bb{v}}_t^z \boxminus v_t^z
 \end{aligned}\right].
\label{Eqn:SELD}
\end{align}
\ES 
where $\hat{v}_t^x$, $\hat{v}_t^z$, $\hat{\bb{v}}_t^x$, $\hat{\bb{v}}_t^z$ denote the output \emph{a posteriori} variances of $\hat{\psi}_t(\bm{\bb{z}}_t)$, $\hat{\phi}_t(\bm{\bb{x}}_t)$, $\hat{\gamma}_t^z({\bm{z}}_t, {\bm{x}}_t)$, and $\hat{\gamma}_t^x({\bm{z}}_t, {\bm{x}}_t)$, i.e.,
\BS \label{Eqn:post_vari}
\begin{align} \nonumber
    \hat{v}_t^x &\overset{\rm a.s.}{=} \hat{\phi}_{\mr{SE}}(\bb{v}_t^x)=\tfrac{1}{N}\mr{E}\{||\hat{\phi}_t(\bf{x}+\sqrt{\bb{v}_t^x}\bf{\eta}_t^x)-\bf{x}||^2\},\\  \nonumber
    \hat{v}_t^z &\overset{\rm a.s.}{=}\hat{\psi}_{\mr{SE}}(\bb{v}_t^z)=\tfrac{1}{M}\mr{E}\{||\hat{\psi}_t(\bf{z}+\sqrt{\bb{v}_t^z}\bf{\eta}_t^z)-\bf{z}||^2\},\\ \nonumber 
    \!\hat{\bb{v}}_t^x &\overset{\rm a.s.}{\!\!=\!\!}\hat{\gamma}_{\mr{SE}}^x(v_t^x, v_t^z)\!\!=\!\!\tfrac{1}{N}\mr{E}\{||\hat{\gamma}^x_t(\bf{x}\!\!+\!\!\sqrt{{v}_t^x}\bf{\eta}_t^x, \bf{z}\!\!+\!\!\sqrt{v_t^z}\bf{\eta}_t^z)\!\!-\!\!\bf{x}||^2\},\\ \nonumber 
    \hat{\bb{v}}_t^z &\overset{\rm a.s.}{\!\!=\!\!} \hat{\gamma}_{\mr{SE}}^z(v_t^x, v_t^z)\!\!=\!\!\tfrac{1}{M}\mr{E}\{||\hat{\gamma}^z_t(\bf{x}\!\!+\!\!\sqrt{{v}_t^x}\bf{\eta}_t^x, \bf{z}\!\!+\!\!\sqrt{v_t^z}\bf{\eta}_t^z)\!\!-\!\!\bf{z}||^2\},
\end{align}
\ES
where $\bf{\eta}_t^x \sim \mathcal{CN}(\bf{0}, \bf{I})$, $\bf{\eta}_t^z \sim \mathcal{CN}(\bf{0}, \bf{I})$, $\bf{\eta}_t^x$ is independent of $\bf{\eta}_t^z$, and $\bf{\eta}_t^x$ and $\bf{\eta}_t^z$ are independent of $\bf{x}$ and $\bf{z}$, respectively. Fig.~\ref{Fig:GOAMP_SE} gives a graphical illustration of the SE in \eqref{Eqn:SE_GOAMP}.

% \emph{Note:} The LDPC decoder $\hat{\phi}_t(\cdot)$ is proved to be Lipschitz-continuous in \cite[Appendix B]{ebert2023sparse}, indicating that the SE of GOAMP/GVAMP based on LDPC decoding holds. As a result, we design a kind of LDPC code based on the SE in numerical results.  Although there is no strict proof for other types of FEC codes, we conjecture that $\hat{\phi}_t(\cdot)$ is also Lipschitz-continuous for the majority of FEC codes (e.g., Turbo code, Polar code, etc.).

\vspace{-0.1cm}
\subsection{Variational State Evolution (VSE)}\label{Sec:VSE} %\vspace{-0.2cm}
\begin{figure}[!t] \vspace{-0.7cm}
\centering 
\subfigure[II-GOAMP/GVAMP receiver: {ELD $\bar{\gamma}_t$ and ${\text{NLD}}_x$ $\hat{\phi}_{t}$, where $\bar{\gamma}_{t}$ includes the orthogonal operations (Orth) and the adequate internal iterations between $\hat{\psi}_{\tau}$ and $\hat{\gamma}_{t}$.}]{
\includegraphics[width=0.8\columnwidth] {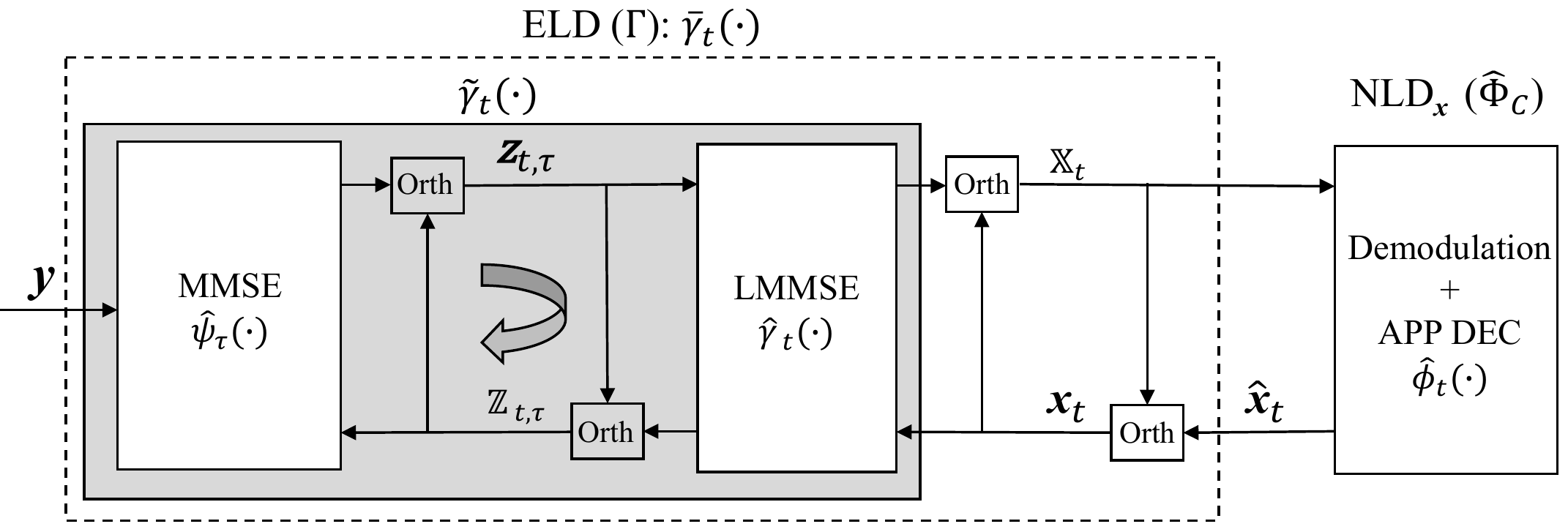}\label{Fig:Var_GOAMP}\vspace{-0.1cm}
}
\vspace{-0.1cm}\centering \vspace{-0.1cm}
\subfigure[Variational transfer functions: $\bar{\gamma}_{\text{SE}}$, $\tilde{\gamma}_{\text{SE}}$, $\hat{\psi}_{\text{SE}}$, $\hat{\gamma}_{\text{SE}}$, and $\hat{\phi}_{\text{SE}}$ denote MSE functions of $\bar{\gamma}_t$, $\tilde{\gamma}_t$, $\hat{\psi}_{\tau}$, $\hat{\gamma}_{t}$, and $\hat{\phi}_{t}$, respectively.
]{\includegraphics[width=0.8\columnwidth]{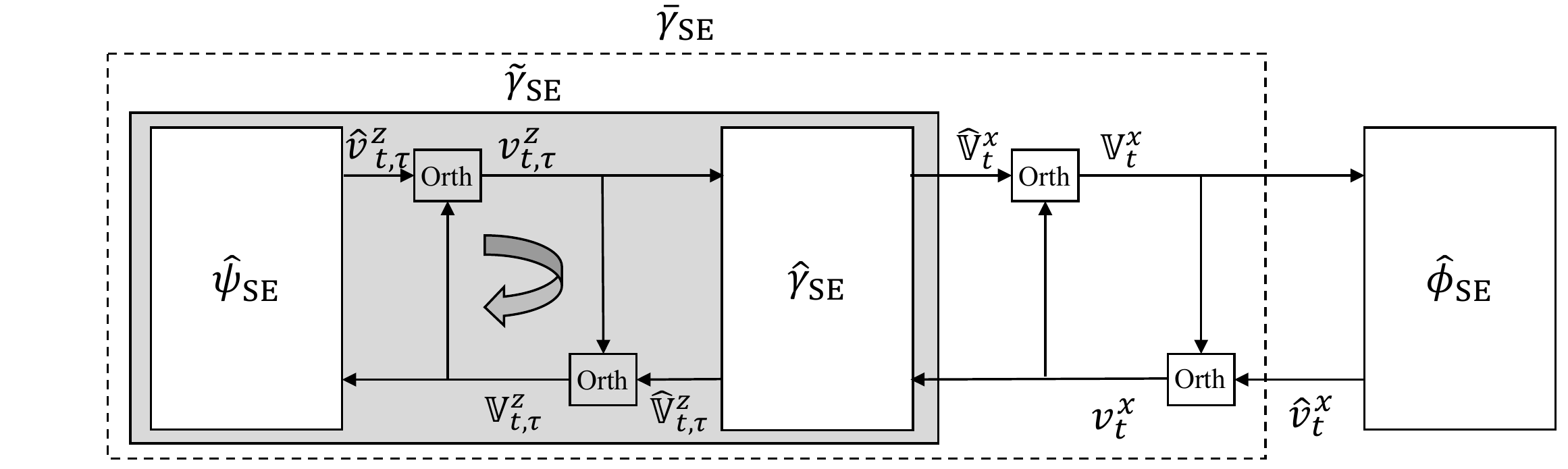}\label{Fig:SE_GLM}
} \vspace{-0.3cm}
\caption{Graphical illustrations for (a) II-GOAMP/GVAMP receiver and (b) the variational transfer functions.} \label{Fig:var_GOAMP_SE} 
\end{figure}

Note that the SE of GOAMP/GVAMP is multi-layer and involves DIDO transfer functions, making it difficult to directly apply the achievable rate analysis of OAMP based on single-layer SLS\cite{LeiOptOAMP}. To address this difficulty, we study an alternative inner-iterative (II) GOAMP/GVAMP, as shown in Fig.~\ref{Fig:Var_GOAMP}, where $\text{NLD}_z$, LD, and the orthogonalization procedures are combined to form an enhanced LD (ELD). Specifically, for given input $\hat{\bf{x}}_t$ from $\text{NLD}_{x}$, the internal iteration  between $\hat{\psi}_t(\cdot)$ and ${\hat{\gamma}}_t(\cdot)$ is executed in $\bar{\gamma}_t(\cdot)$ until it converges. Based on this, the GLS is converted to an SLS composed of $\bar{\gamma}_t(\cdot)$ and $\hat{\phi}_t(\cdot)$, while in SE, the original DIDO transfer functions are transformed into SISO transfer functions, as shown in Fig.~\ref{Fig:SE_GLM}.

\emph{II-GOAMP/GVAMP Receiver}: Starting with $t=1$ and $\bf{\bb{x}}_1=0$,
\BE\label{Eqn:II-GOAMP}
\text{NLD:}
\;\;{{\hat{\bm{x}}_t}} = \hat{\phi}_t(\bf{\bb{x}}_t), \;\;
\text{LD:}\;\;{\bf{\bb{x}}_{t+1}} = \bar{\gamma}_t(\hat{\bm{x}}_t, {\bf{\bb{x}}_{t}}),
\EE
where $\bar{\gamma}_t(\hat{\bm{x}}_t, {\bf{\bb{x}}_{t}})$ includes the orthogonal operations and the adequate internal iterations between $\hat{\gamma}_{t}({\bm{x}}_t, \bm{{z}}_{t,\tau})$ and $\hat{\psi}_{\tau}(\bm{\bb{z}}_{t,\tau})$, $\tau$ denotes the inner iteration index in $\bar{\gamma}_t$, and $t$ the outer iteration index between $\bar{\gamma}_t$ and $\hat{\phi}_t$. We will not provide the specific expression of $\bar{\gamma}_t$, which is not relevant to the discussions in this paper. However, we will discuss in detail the SE of II-GOAMP/GVAMP, which serves as a bridge to the achievable rate analysis and code design of the original GOAMP/GVAMP. The SE of II-GOAMP/GVAMP is given by
\vspace{-0.4cm}
\BS\label{Eqn:DIDOSE_IIGOAMP}
\begin{align}
\!\!\!\!\text{NLD:}&\;
\hat{v}_t^x = \hat{\phi}_{\mr{SE}}({\bb{v}}_{t}^x), \label{Eqn:DIDOSE_IINLD}\\
\!\!\!\!\text{LD:} & \;
{\bb{v}}_{{t+1}}^x \!\!= \!\!\bar{\gamma}_{\mr{SE}
}(\hat{v}_t^x, {\bb{v}}_{{t}}^x)\!\!= \!\!\tilde{\gamma}_{\mr{SE}}(\hat{v}_t^x \boxminus \bb{v}_t^x)\boxminus ( \hat{v}_t^x \boxminus \bb{v}_t^x ), 
\label{Eqn:DIDOSE_IILD}
\end{align}
\ES
where $\tilde{\gamma}_{\mr{SE}}(\cdot)$ involves adequate inner iterations between MSE transfer functions $\hat{\psi}_{\mr{SE}}$ (for $\hat{\psi}_{\tau}$) and $\hat{\gamma}_{\mr{SE}}$ (for $\hat{\gamma}_{t}$), i.e.,
% \BS
\BE\label{Eqn:II_gamma_bar}
{\hat{\bb{v}}}_{{t+1}}^x = \tilde{\gamma}_{\mr{SE}}(v_t^x)=\hat{\gamma}_{\mr{SE}}^{x}(v_t^x, v^z_{t,*}),
\EE
with $v_t^x= \hat{v}_t^x \boxminus \bb{v}_t^x$, and  $v^z_{t,*}$ is the fixed point of the  inner iteration between  
% {\setlength\abovedisplayskip{2pt}
% \setlength\belowdisplayskip{2pt}
\BE \nonumber
{\bb{v}}_{t,\tau}^z ={\hat{\bb{v}}}_{t,\tau}^z\boxminus v_{t,\tau}^z,  \;\; 
v_{t,\tau+1}^z =\hat{v}^z_{t,\tau}\boxminus \bb{v}_{t,\tau}^z.
\EE
% }
% \ES
Precisely, $v^z_{t,*}$ is a function of $v_t^x$ and can be solved by 
{\setlength\abovedisplayskip{4pt}
\setlength\belowdisplayskip{4pt}
\BE\label{Eqn:II_local_FP}
{\bb{v}}_{t,*}^z =\hat{\gamma}_{\mr{SE}}^z(v_t^x,v_{t,*}^z)\boxminus v_{t,*}^z, \;\; 
v_{t,*}^z  =\hat{\psi}_{\mr{SE}}(\bb{v}_{t,*}^z)\boxminus \bb{v}_{t,*}^z.
\EE
}

Note that the transfer function $\bar{\gamma}_{\mr{SE}}(\cdot)$ in \eqref{Eqn:DIDOSE_IILD} is a dual-input-single-output (DISO) function that incorporates the additional $\bb{v}_t^x$ from the previous iteration, although the new SE in \eqref{Eqn:DIDOSE_IIGOAMP} is simpler than the original SE in \eqref{Eqn:SE_GOAMP}. As a result, it is still difficult to obtain the achievable rate of II-GOAMP/GVAMP using the I-LMMSE lemma. This problem can be solved by converting the DISO function $\bar{\gamma}_{\mr{SE}}(\cdot)$ into a SISO function $\breve{\gamma}_{\mr{SE}}(\cdot)$ by replacing $\bb{v}_t^x$ with $\bb{v}_{t+1}^x$, which does not change the SE fixed points of II-GOAMP/GVAMP in \eqref{Eqn:DIDOSE_IIGOAMP} \cite{LeiOptOAMP}. See \cite[Lemma 4]{LeiOptOAMP} for the details. Then, based on \eqref{Eqn:DIDOSE_IIGOAMP}, we can obtain the following variational SISO transfer functions.  
\BS\label{Eqn:SE_IIGOAMP}
\begin{align}
\text{NLD:}&\;\;
\hat{v}_t^x = \hat{\phi}_{\mr{SE}}({\bb{v}}_{t}^x), \label{Eqn:SEIINLD}\\
\text{LD:}& \;\;
{\bb{v}}_{{t+1}}^x = \breve{\gamma}_{\mr{SE}}(\hat{v}_t^x)=\hat{v}_t^x \boxminus \tilde{\gamma}_{\mr{SE}}^{-1}(\hat{v}_t^x), 
\label{Eqn:SEIILD}
\end{align}
where $\tilde{\gamma}_{\mr{SE}}^{-1}(\cdot)$ is the inverse of $\tilde{\gamma}_{\mr{SE}}(\cdot)$.
\ES

The following lemma indicates that the converged MSE of II-GOAMP/GVAMP is the same as that of the original GOAMP/GVAMP.

\begin{lemma}[Equivalence of GOAMP/GVAMP and II-GOAMP/GVAMP]\label{Lem:Covge_consis}
GOAMP/GVAMP and II-GOAMP/ GVAMP have the same SE fixed points given by
\BS\label{Eqn:SEFP_GOAMP}
\begin{align}
\text{NLD:}\quad &
\left[ \begin{aligned}  v_*^x \\ v_*^z  \end{aligned} \right] = 
\left[ \begin{aligned} {{\phi}_{\mr{SE}}({\bb{v}}_*^x)} \\ {{\psi}_{\mr{SE}}({\bb{v}}_*^z)} \end{aligned} \right] = \left[ \begin{aligned}  
\hat{\phi}_{\mr{SE}}(\bb{v}_*^x)\boxminus{{ \bb{v}}_*^x}\\
\hat{\psi}_{\mr{SE}}({\bb{v}}_*^z)\boxminus {{\bb{v}}_*^z}
\end{aligned} \right], \label{Eqn:SE_FP_NLD}\\
\text{LD:}\quad  & \left[ \begin{aligned} {\bb{v}}_{*}^x\\ {\bb{v}}_{*}^z  \end{aligned} \right] = {\gamma}_{\mr{SE}}(v_{*}^x, v_{*}^z) =
\left[ \begin{aligned}
&\hat{\gamma}_{\mr{SE}}^x(v_*^x, v_*^z)\boxminus v_*^x \\
&\hat{\gamma}_{\mr{SE}}^z(v_*^x, v_*^z)\boxminus v_*^z  
 \end{aligned}\right],
\label{Eqn:SE_FP_LD}
\end{align}
\ES
where $\hat{\phi}_{\mr{SE}}$, $\hat{\psi}_{\mr{SE}}$, $\hat{\gamma}_{\mr{SE}}^x$, and $\hat{\gamma}_{\mr{SE}}^z$ are given in \eqref{Eqn:post_vari}.
Moreover, the SE fixed points represent the converged MSEs of GOAMP/ GVAMP and II-GOAMP/GVAMP. Therefore, their converged MSEs are the same.
\end{lemma}

It should be emphasized that the convergence speed and final estimation (not the converged MSE) of II-GOAMP/GVAMP may differ from those of GOAMP/GVAMP.  Following Lemma \ref{Lem:Covge_consis}, the achievable rate analysis and optimal code design of II-GOAMP/GOAMP and GOAMP/GVAMP are the same since they are only determined by the converged MSE. Consequently, the achievable rate  of GOAMP/GVAMP can be analyzed through the variational SISO transfer functions \eqref{Eqn:SE_IIGOAMP}. Due to this fact, we will not distinguish II-GOAMP/GVAMP and GOAMP/GVAMP, and refer to both as GOAMP/GVAMP. 

\begin{figure}[t]\vspace{-0.7cm}
\centering 
\includegraphics[width=0.65\columnwidth]{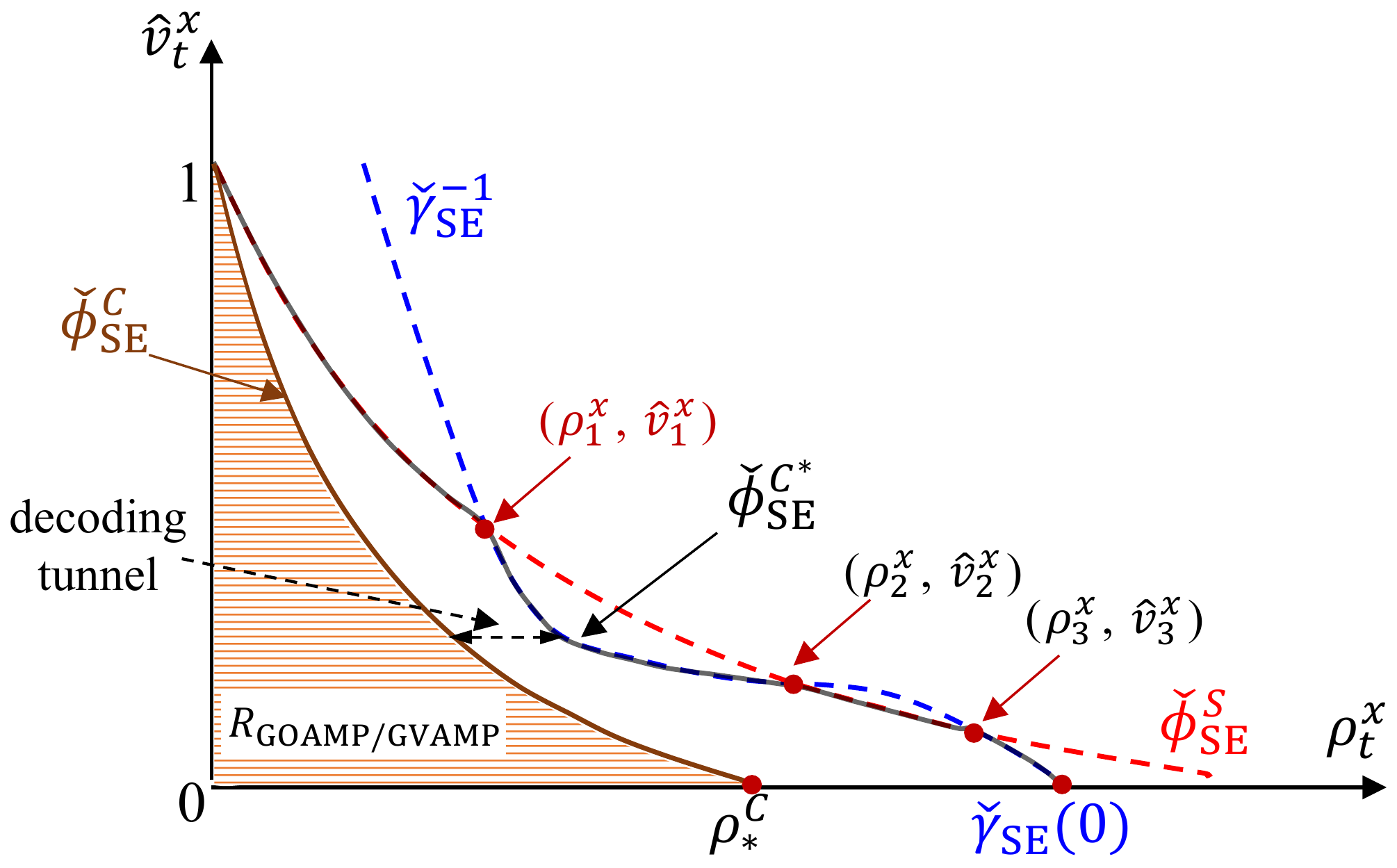}\vspace{-0.3cm}
\caption{Graphical illustration for SE of GOAMP/GVAMP, where $\check{\gamma}_{\mr{SE}}^{-1}(\cdot)$ is the inverse of $\check{\gamma}_{\mr{SE}}(\cdot)$, ${\check{\phi}}_{\mr{SE}}(\cdot)$ is the MMSE function of $\text{NLD}_x$, $\phi_{\mr{SE}}^{\mathcal{S}}(\cdot)$ and $\phi_{\mr{SE}}^{\mathcal{C}}(\cdot)$ is the MMSE functions of demodulator and decoder, and $\phi_{\mr{SE}}^{\mathcal{C}^*}(\cdot)$ is the MMSE function of optimal code. $\rho_{\mathcal{C}}^*$ and $\check{\gamma}_{\text{SE}}(0)$ are the intersections of $\phi_{\mr{SE}}^{\mathcal{C}}(\cdot)$ and $\check{\gamma}^{-1}_{\mr{SE}}(\cdot)$  with the horizontal axis, respectively.} \label{Fig:GSE_Area}\vspace{-0.3cm}
\end{figure}
\subsection{Achievable Rate Analysis and Coding Principle}\label{Sec:achiev_rate}
 Based on Lemma \ref{Lem:Covge_consis}, we can rigorously analyze the achievable rate analysis and optimal code design for GOAMP/GVAMP based on the SISO VSE in \eqref{Eqn:SE_IIGOAMP}. Let $\rho_t^x = 1/{\scriptstyle \mathbb{V}}_{t}^x$. We rewrite the VSE in \eqref{Eqn:SE_IIGOAMP} as
\BE\label{Eqn:SE_IIGOAMP2}
\text{NLD:}\;\;
\hat{v}_t^x = \check{\phi}_{\mr{SE}}(\rho_t^x), \quad
\text{LD:} \;\;
 \rho_{t+1}^x = \check{\gamma}_{\mr{SE}}(\hat{v}_t^x).
\EE

 Due to coding gain, the decoding transfer function $\check{\phi}_{\mr{SE}}^{\mathcal{C}}(\cdot)$ is upper bounded by the demodulation transfer function $\check{\phi}_{\mr{SE}}^{\mathcal{S}}(\cdot)$ \cite{LeiTIT2021,LeiOptOAMP}: 
\BE\label{Eqn:code_gain}
\check{\phi}_{\mr{SE}}^{\mathcal{C}}(\rho_t^x)< \check{\phi}_{\mr{SE}}^{\mathcal{S}}(\rho_t^x),\quad {\mr{for}} \;\; 0\le \rho_t^x \le snr.
\EE
As shown in Fig.~\ref{Fig:GSE_Area}, assume that there are multiple fixed points between $\check{\gamma}^{-1}_{\mr{SE}}(\cdot)$ and $\check{\phi}_{\mr{SE}}^{\mathcal{S}}(\cdot)$ in SE of GOAMP/GVAMP, i.e., $(\rho_1^x, {\hat{v}_1^x})$, $(\rho_2^x, {\hat{v}_2^x})$ and $(\rho_3^x, {\hat{v}_3^x})$, where $\check{\gamma}^{-1}_{\mr{SE}}(\cdot)$ is the inverse of $\check{\gamma}_{\mr{SE}}(\cdot)$. It is obvious that the iterative process of the SE transfer functions will stop at the first fixed point $(\rho_1^x, {\hat{v}_1^x})$. Since ${\hat{v}_1^x}>0$, the converged performance of GOAMP/GVAMP is not error-free. Therefore, to achieve error-free performance, a kind of proper FEC code should be designed to ensure that a decoding tunnel between $\check{\phi}_{\mr{SE}}^{\mathcal{C}}(\cdot)$ and $\check{\gamma}^{-1}_{\mr{SE}}(\cdot)$ is available for successful decoding \cite{LeiTIT2021, LeiOptOAMP}. That is, 
\BE \label{Eqn:SE_coded}
\check{\phi}_{\mr{SE}}^C(\rho_t^x)<\check{\gamma}^{-1}_{\mr{SE}}(\rho_t^x), \quad {\mr{for}} \;\; 0 \le \rho_t^x < [\check{\gamma}_{\mr{SE}}(0)].
\EE
Then, based on the I-MMSE lemma~\cite{GuoTIT2005}, we give the achievable rate of GOAMP/GVAMP in GLS in the following lemma.

\begin{figure}[!t]\vspace{-0.4cm}
\centering 
\includegraphics[width=0.56\columnwidth]{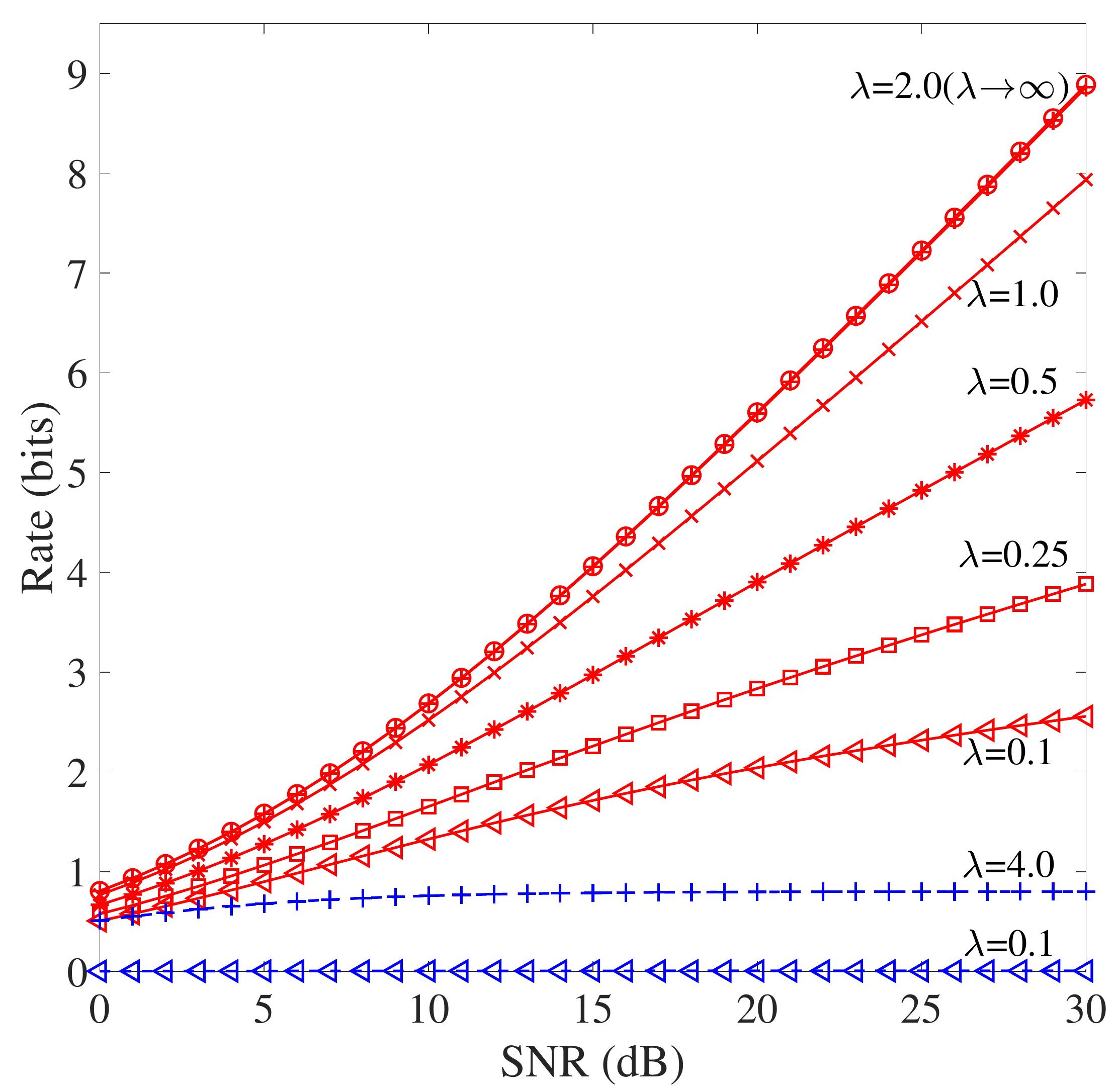}\vspace{-0.35cm}
\caption{Maximum achievable rates of GOAMP/GVAMP, the linearized approximate clipping model based on MRC receiver (Linear-Clipping-MRC)\cite{ShansuoWCL2019,Tong2010_clip} with $N=500$, $\delta=1$, Gaussian signaling, channel condition number $\kappa=10$, and clipping thresholds $\lambda=\{0.1, 0.25, 0.5, 1, 2, \infty\}$.} \label{Fig:Rate_of_clip}\vspace{-0.25cm}
\end{figure}
\begin{lemma}[Achievable Rate of GOAMP/GVAMP]\label{Lem:rate_II_GOAMP} 
The achievable rate of GOAMP/ GVAMP with fixed $\check{\phi}^{C}_{\mr{SE}}(\cdot)$ is 
\BE \label{Eqn:rate_II_GOAMP}
\begin{aligned}
 &R_{\text{GOAMP/GVAMP}} = \int_{0}^{\check{\gamma}_{\mr{SE}}(0)} \check{\phi}^{C}_{\mr{SE}}(\rho_t^x) d \rho_t^x,   \\
 &\begin{array}{l@{\quad}l}
 {\rm s.t.} &  \check{\phi}_{\mr{SE}}^{\mathcal{C}}(\rho_t^x)< \check{\phi}_{\mr{SE}}^{{\mathcal{C}}^*}(\rho_t^x), \quad {\mr{for}} \;\; 0\le \rho_t^x \le \check{\gamma}_{\mr{SE}}(0),
 \end{array}
\end{aligned}
\EE
\end{lemma}
 where $\check{\phi}_{\mr{SE}}^{{\mathcal{C}}^*}(\rho_t^x)= {\mr {min}}\{\check{\phi}_{\mr{SE}}^{\mathcal{S}}(\rho_t^x),\check{\gamma}^{-1}_{\mr{SE}}(\rho_t^x)\}$.

 Based on Lemma \ref{Lem:rate_II_GOAMP}, we obtain the optimal coding principle to maximize the achievable rate of GOAMP/GVAMP in the following lemma.

\begin{lemma}[Optimal Code Design]\label{lem:error_free}
The optimal coding principle is to match the MMSE decoding transfer function with $\check{\phi}_{\mr{SE}}^{{\mathcal{C}}^*}(\rho_t^x)$, i.e.,
\BE\label{Eqn:error_free}
\check{\phi}_{\mr{SE}}^{\mathcal{C}}(\rho_t^x) \rightarrow \check{\phi}_{\mr{SE}}^{{\mathcal{C}}^*}(\rho_t^x), \quad {\mr{for}} \;\; 0\le \rho_t^x \le \check{\gamma}_{\mr{SE}}(0),
\EE
which achieves the maximum achievable rate while ensuring error-free performance.
\end{lemma}  

The following theorem follows straightforwardly Lemma \ref{Lem:rate_II_GOAMP} and Lemma \ref{lem:error_free}. It presents the maximum achievable rate of GOAMP/GVAMP that allows multiple fixed points between ${\check{\phi}}^{\mathcal{S}}_{\mr{SE}}(\cdot)$ and $\check{\gamma}^{-1}_{\mr{SE}}(\cdot)$. Note that the results in \cite{LeiOptOAMP} are dependent on a unique fixed point assumption for SE, which does not always hold.  

 \begin{theorem}[Maximum Achievable Rate]\label{them:capacity}
 The maximum achievable rate of GOAMP/GVAMP is
\BE\label{Eqn:con_capacity}
R_{\text{GOAMP/GVAMP}}^{\text{max}}\rightarrow \int_0^{\check{\gamma}_{\mr{SE}}(0)}\check{\phi}_{\mr{SE}}^{\mathcal{C}^*}(\rho_t^x) d \rho_t^x.
\EE
\end{theorem}

\section{Numerical Results}
\subsection{System Configuration}\label{Sec:sysm_config}
% Note that $\bf{A}$ is unitarily-invariant and let SVD of $\bm{A}$ be $\bm{A}=\bm{U}\bm{\Lambda}\bm{V}$, where $\bm{\Lambda}$ is a rectangular diagonal matrix, $\{\bm{U}, \bm{\Lambda}, \bm{V}\}$ are independent, and $\bm{U}$ and $\bm{V}$ are Haar distributed (i.e., uniformly distributed over all unitary matrices) \cite{RandomWire}. 
Assume that the channel matrix $\bf{A}\in\mathbb{C}^{M \times N}$ is  unitarily-invariant and fixed during the transmission, where $N=500$ and the condition number of $\bf{A}$ is $\kappa=10$. Let the SVD of $\bf{A}$ be $\bf{A}=\bf{U}\bf{\Lambda}\bf{V}^{H}$,  where $\bm{\Lambda}$ is a rectangular diagonal matrix, $\{\bm{U}, \bm{\Lambda}, \bm{V}\}$ are independent, and $\bm{U}$ and $\bm{V}$ are Haar distributed (i.e., uniformly distributed over all unitary matrices) \cite{RandomWire}. To reduce the calculation complexity of matrix multiplication, we approximate two large random unitary matrices by $\bf{U}=\bf{F}_1\bf{\Pi}_1$ and $\bf{V}^{H}=\bf{\Pi}_2\bf{F}_2$, where $\bf{\Pi}_1$, $\bf{\Pi}_2$ are random permutation matrices and $\bf{F}_1$, $\bf{F}_2$ are discrete Fourier transform (DFT) matrices with dimensions $M$ and $N$ \cite{GMAMP2022}. The singular values $\{d_i\}$ in $\bf{\Lambda}$ are set as \cite{Vila2015ICASSP}: $d_i/d_{i+1}=\kappa^{1/\mathcal{T}}$, $i=1, ..., \mathcal{T}-1$ and $\sum_{i=1}^{\mathcal{T}} d_i^2 =\mathcal{J}$, where $\mathcal{T}={\mr{min}}\{M, N\}, \mathcal{J}={\mr{max}}\{M, N\}$.

\subsection{Achievable Rate of GOAMP/GVAMP with Clipping}
Clipping is commonly applied to reduce the PAPR in OFDM systems\cite{clip_cl}. 
Here, we provide the maximum achievable rate of GOAMP/GVAMP with clipping under different clipping thresholds $\lambda$ based on Theorem~\ref{them:capacity}. As shown in Fig.~\ref{Fig:Rate_of_clip}, for a fixed $\lambda$, the achievable rate of GOAMP/GVAMP increases monotonically with SNR. When $\lambda < 2$, the achievable rate increases as the $\lambda$ increases. For $\lambda\ge 2$, the achievable rates will no longer increase, i.e., the achievable rate curves with $\lambda=2$ and $\lambda\rightarrow\infty$ overlap. Because the clip threshold $\lambda\ge 2$ already exceeds the range of noisy observation values, it is not the main factor limiting the achievable rate. Moreover, as shown in Fig.~\ref{Fig:Rate_of_clip}, the achievable rate of conventional MRC receiver based on the linearized model is much lower that of the proposed GOAMP/GVAMP. This is due to the linearized model's inherent higher rate loss, and secondly, interference between signals in the MRC receiver is not well suppressed.

\begin{figure}\vspace{-0.35cm}
\centering 
\includegraphics[width=0.95\columnwidth]{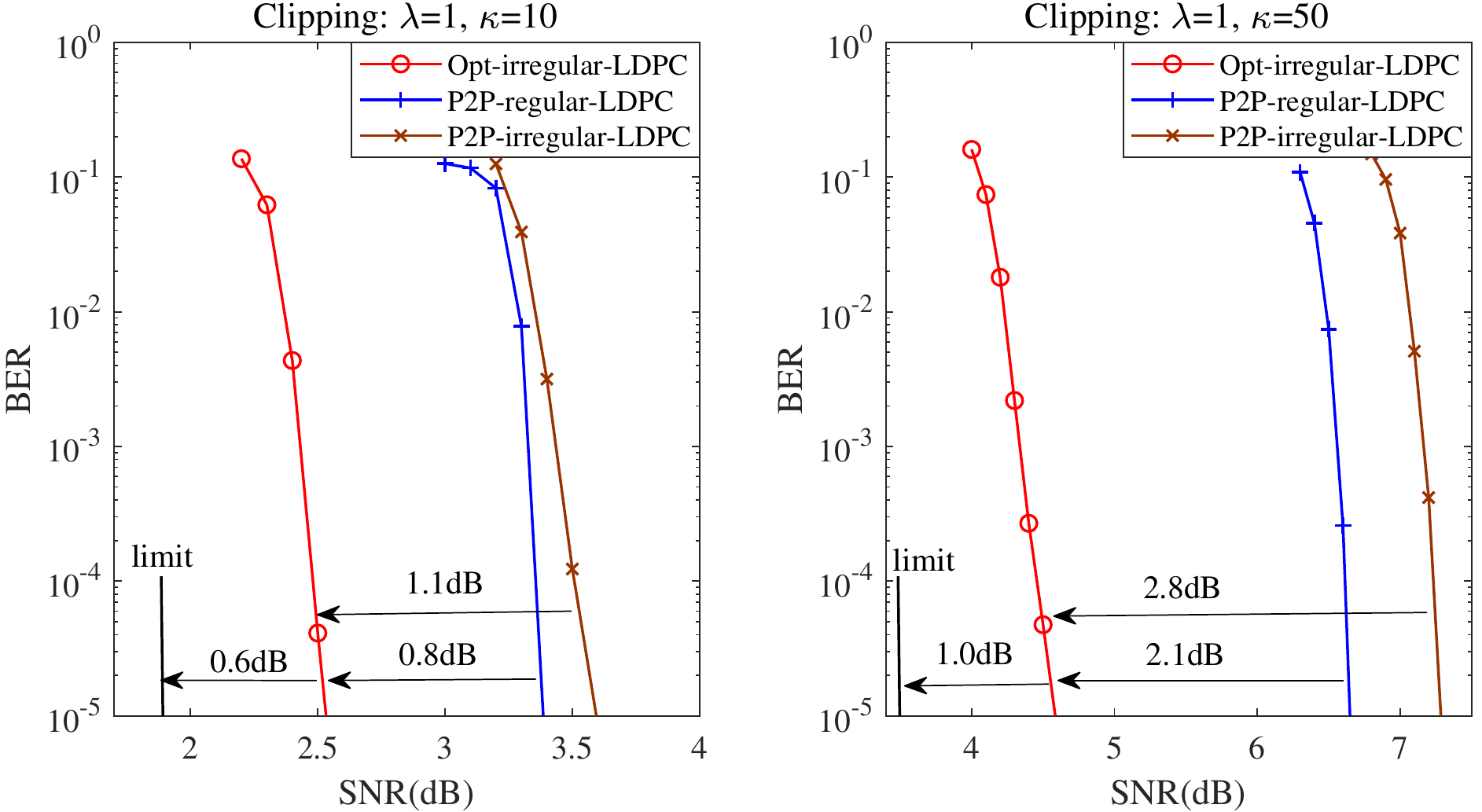}\vspace{-0.3cm}
\caption{BER performances of GOAMP/GVAMP with the optimized LDPC codes, the $R_{\rm{LDPC}}=0.5$ P2P regular (3,6) and P2P capacity-approaching irregular LDPC codes \cite{Richardson2001}. "limit" denotes the associated information-theoretic limit of GOAMP/GVAMP.} \label{Fig:BER_compare}\vspace{-0.15cm}
\end{figure}

\subsection{BER Simulations and Comparisons}
Note that the principle of optimal code design in Lemma~\ref{lem:error_free} is applicable for arbitrary input distributions. Thus, we consider the practical LDPC code design for QPSK signaling with target rate $R_{\mr{sum}}=NR_{\mr{LDPC}}{\mr{log}}_{2}|\mathcal{S}_{\mr{QPSK}}|=500$, where $R_{\mr{LDPC}}$$=0.5$ and $|\mathcal{S}_{\mr{QPSK}}|$$=4$. Fig. \ref{Fig:BER_compare} shows the gaps between BER curves at $10^{-4}$ of the optimized LDPC codes and the corresponding theoretical limits are about within $1.0$~dB, which verifies the near optimality of the proposed LDPC codes. Moreover, compared with P2P regular $(3, 6)$ LDPC codes and well-designed irregular LDPC codes  with $R_{\rm{LDPC}}=0.5$ \cite{Richardson2001}, the optimized LDPC codes with GOAMP/GVAMP can achieve about $0.8 \sim 2.8$~dB performance gains. The above comparison demonstrates that the Bayes optimal GOAMP/GVAMP cannot be guaranteed to achieve error-free performance.

\section*{Conclusion}
This paper focuses on the achievable rate analysis and coding principle of GOAMP/GVAMP for the GLS with unitarily-invariant matrices and arbitrary input distributions. Based on the same convergence performance as the original DIDO transfer functions, the equivalent variational SISO transfer functions are proposed for the analysis of achievable rates and optimal code design. Numerical results demonstrate the achievable rate advantages of the proposed GOAMP/GVAMP and the BER performance gains of the optimized LDPC compared to the existing methods. 

%Moreover, the achievable rate analysis and code design of GOAMP/GVAMP can be further extended to  quantization and more general high-dimensional neural networks\cite{LeiGOAMP}.

% \section*{Acknowledgment}

\clearpage
\bibliographystyle{IEEEtran}
\bibliography{manuscript}

\clearpage
\begin{appendices}
\section{Orthogonality and Asymptotic IID Gaussianity}\label{APP:IIDG}%}
Define error vectors as
\BS \label{Eqn:est_error}
\begin{align}
    \bf{\xi}^x_t = \bf{\bb{x}}_t-\bf{x}, \qquad \bf{\xi}_t^z = \bf{z}-\bf{\bb{z}}_t,\\
    \bf{\epsilon}_t^x = \bf{x}-\bf{x}_t, \qquad \bf{\epsilon}_t^z = \bf{z}_t-\bf{z},
\end{align}%\vspace{-0.4cm}
\ES
with zero mean and variances:
%\vspace{-0.1cm}
\BS \label{Eqn:est_variance}
\begin{align}
    v_t^x = \left \langle \bf{\xi}_t^x|\bf{\xi}_t^x \right \rangle, \qquad v_t^z = \left \langle \bf{\xi}_t^z|\bf{\xi}_t^z \right \rangle,\\
    {\bb{v}}_{t}^x=\left \langle\bf{\epsilon}_t^x|\bf{\epsilon}_t^x\right \rangle,  \qquad {\bb{v}}_{t}^z=\left \langle\bf{\epsilon}_t^z|\bf{\epsilon}_t^z\right \rangle.
\end{align}
\ES

The following lemma establishes the asymptotic IID Gaussianity of GOAMP/GVAMP based on the unitarily-invariant property of $\bf{A}$~\cite{MontanariSE,Rangan2019TIT, Takeuchi2020} (see also \cite{GMAMP2022} for more details). 

\begin{lemma} [Orthogonality and Asymptotic IID Gaussianity]\label{Lem:Orth_IIDG}
Assume that $\hat{\psi}_t(\cdot)$, $\hat{\gamma}_t(\cdot)$, and $\hat{\phi}_t(\cdot)$ are Lipschitz-continuous estimators. The following orthogonality holds for the iterative process of GOAMP/GVAMP: $\forall t>1$,
\BS \label{Eqn:orth}
\begin{align}
    \left \langle \bf{\xi}_t^x|{\epsilon}_t^x\right \rangle &\overset{\rm a.s.}{=}0, \qquad \left \langle \bf{\xi}_t^z|\bf{\epsilon}_t^z \right \rangle \overset{\rm a.s.}{=} 0,\\
    \left \langle\bf{\epsilon}_{t+1}^x|\bf{\xi}_t^x\right \rangle &\overset{\rm a.s.}{=}0,  \qquad \left \langle\bf{\epsilon}_{t+1}^z|\bf{\xi}_t^z\right \rangle \overset{\rm a.s.}{=}0.
\end{align}
\ES
Then, the following asymptotic IID Gaussianity holds: for $t>1$,
\BS \label{Eqn:AIIG}
\begin{align}
\!\!\!v_t^x &\overset{\rm a.s.}{=}\tfrac{1}{N}\mr{E}\{||\phi_t(\bf{x}+\sqrt{ \bb{v}_t^x}\bf{\eta}_t^x)-\bf{x}||^2\},\\
\!\!\!{\bb{v}}_{t+1}^x &\overset{\rm a.s.}{=} \tfrac{1}{N}\mr{E}\{||\gamma^x_t(\bf{x}+\sqrt{{v}_t^x}\bf{\eta}_t^x, \bf{z}+\sqrt{v_t^z}\bf{\eta}_t^z)-\bf{x}||^2\},\\
\!\!\!{\bb{v}}_{t+1}^z &\overset{\rm a.s.}{=} \tfrac{1}{M}\mr{E}\{||\gamma^z_t(\bf{x}+\sqrt{{v}_t^x}\bf{\eta}_t^x, \bf{z}+\sqrt{v_t^z}\bf{\eta}_t^z)-\bf{z}||^2\},
\end{align}
where $\bf{\eta}_t^x \sim \mathcal{CN}(\bf{0}, \bf{I})$, $\bf{\eta}_t^z \sim \mathcal{CN}(\bf{0}, \bf{I})$, $\bf{\eta}_t^x$ is independent of $\bf{\eta}_t^z$, and $\bf{\eta}_t^x$ and $\bf{\eta}_t^z$ are independent of $\bf{x}$ and $\bf{z}$, respectively.
\ES
\end{lemma}

\emph{Note:} The $\hat{\gamma}_t(\cdot)$ has been proven to be Lipschitz-continuous in \cite{Takeuchi2020}. $\hat{\psi}_t(\cdot)$ is naturally Lipschitz-continuous because it is demodulated symbol-by-symbol in this paper. Meanwhile, the LDPC decoder $\hat{\phi}_t(\cdot)$ is proved to be Lipschitz-continuous in \cite[Appendix B]{ebert2023sparse}, indicating that the SE of GOAMP/GVAMP based on LDPC decoding holds. As a result, we design a kind of LDPC code based on the SE in numerical results. Although no rigorous proof exists for other types of FEC codes, we conjecture  it is possible to demonstrate that $\hat{\phi}_t(\cdot)$ is Lipschitz-continuous.  

\section{Proof of Lemma~\ref{Lem:Covge_consis}}\label{APP:sameFP}%}
Let $(v_*^z, {\bb{v}}_*^z, v_*^x, {\bb{v}}_*^x)$ be the SE fixed point of GOAMP/GVAMP. Then the fixed-point equation \eqref{Eqn:SEFP_GOAMP2} can be obtained directly from \eqref{Eqn:SE_GOAMP} and \eqref{Eqn:post_vari}. 
\BS\label{Eqn:SEFP_GOAMP2}
\begin{align}
\text{NLD:}\quad &
\left[ \begin{aligned}  v_*^x \\ v_*^z  \end{aligned} \right] = 
\left[ \begin{aligned} {{\phi}_{\mr{SE}}({\bb{v}}_*^x)} \\ {{\psi}_{\mr{SE}}({\bb{v}}_*^z)} \end{aligned} \right] = \left[ \begin{aligned}  
\hat{\phi}_{\mr{SE}}(\bb{v}_*^x)\boxminus{{ \bb{v}}_*^x}\\
\hat{\psi}_{\mr{SE}}({\bb{v}}_*^z)\boxminus {{\bb{v}}_*^z}
\end{aligned} \right], \label{Eqn:SE_FP_NLD2}\\
\text{LD:}\quad  & \left[ \begin{aligned} {\bb{v}}_{*}^x\\ {\bb{v}}_{*}^z  \end{aligned} \right] = {\gamma}_{\mr{SE}}(v_{*}^x, v_{*}^z) =
\left[ \begin{aligned}
&\hat{\gamma}_{\mr{SE}}^x(v_*^x, v_*^z)\boxminus v_*^x \\
&\hat{\gamma}_{\mr{SE}}^z(v_*^x, v_*^z)\boxminus v_*^z  
 \end{aligned}\right].
\label{Eqn:SE_FP_LD2}
\end{align}
\ES

Let $(\hat{v}_*^x, {\bb{v}}_{*}^x)$ be the SE fixed point of II-GOAMP/GVAMP. Based on \eqref{Eqn:SE_IIGOAMP} and \eqref{Eqn:post_vari}, we have the following SE fixed-point equation for II-GOAMP/GVAMP. 
\BS\label{Eqn:SE_FP_IIGOAMP}
\begin{align}
\text{NLD:}&\;\;
\hat{v}_*^x = \hat{\phi}_{\mr{SE}}({\bb{v}}_{*}^x), \label{Eqn:SE_FP_IINLD}\\
\text{LD:}& \;\;
{\bb{v}}_{{*}}^x = \bar{\gamma}_{
\mr{SE}}(\hat{v}_*^x)=\hat{v}_*^x \boxminus \tilde{\gamma}_{\mr{SE}}^{-1}(\hat{v}_*^x), 
\label{Eqn:SE_FP_IILD}
\end{align}
\ES 
where $\tilde{\gamma}_{\mr{SE}}^{-1}(\cdot)$ is the inverse of $\tilde{\gamma}_{\mr{SE}}(\cdot)$, which is given in  \eqref{Eqn:II_gamma_bar} and \eqref{Eqn:II_local_FP}, i.e.,
\BS\label{Eqn:II_local_FP2}
\begin{align} \label{Eqn:SE_pof1}
&  \tilde{\gamma}_{\mr{SE}}(v_*^x)=\hat{\gamma}_{\mr{SE}}^{x}(v_*^x, v^z_{*,*}),\\ \label{Eqn:SE_pof4}
&{\bb{v}}_{*,*}^z= \hat{\gamma}_{\mr{SE}}^z(v_*^x, v_{*,*}^z)\boxminus (v_{*,*}^z),\\ \label{Eqn:SE_pof5}
&v_{*,*}^z= \hat{\psi}_{\mr{SE}}({\bb{v}}_{*,*}^z)\boxminus ({\bb{v}}_{*,*}^z).
\end{align} 
\ES
Substituting $ \hat{v}_*^x = [(v_*^x)^{-1} + (\bb{v}^x_*)^{-1}]^{-1}$ into \eqref{Eqn:SE_FP_IINLD}, we have 
\BE\label{Enq:SE_pof2}
v_*^x = \hat{\phi}_{\mr{SE}}(\bb{v}_*^x)  \boxminus \bb{v}_*^x.
\EE
Similarly, substituting \eqref{Eqn:SE_pof1} and $ \hat{v}_*^x = [(v_*^x)^{-1} + (\bb{v}^x_*)^{-1}]^{-1}$ into \eqref{Eqn:SE_FP_IILD}, we have 
\BE\label{Eqn:SE_pof3}
\bb{v}_*^x = \hat{\gamma}_{\mr{SE}}^x(v_*^x, v_{*,*}^z) \boxminus v_*^x.
\EE
As a result, the SE fixed-point equations of II-GOAMP/GVAMP in \eqref{Eqn:SE_FP_IIGOAMP} are converted to \eqref{Eqn:SE_pof4}, \eqref{Eqn:SE_pof5}, \eqref{Enq:SE_pof2} and \eqref{Eqn:SE_pof3}. By replacing  $v_{*,*}^z$ with $v_*^z$ and ${\bb{v}}_{*,*}^z$ with ${\bb{v}}_{*}^z$, we can see that  \eqref{Enq:SE_pof2} and \eqref{Eqn:SE_pof5} are the same as the LD fixed-point equations in  \eqref{Eqn:SE_FP_NLD2}, and \eqref{Eqn:SE_pof3} and \eqref{Eqn:SE_pof4} are the same as the NLD fixed-point equations in  \eqref{Eqn:SE_FP_LD2}. This indicates that GOAMP/GVAMP and II-GOAMP/GVAMP have the same SE fixed-point equations. That is,  GOAMP/GVAMP and II-GOAMP/GVAMP converge to the same MSE. Hence, we complete the proof of Lemma \ref{Lem:Covge_consis}.

\section{De-clipping}\label{APP:Declip}%}
Clipping is commonly applied to reduce the PAPR in OFDM systems\cite{clip_cl}. Let $z_m$ and $n_m$ be the $m$-th elements of $\bf{z}$ and $\bf{n}$ in \eqref{Eqn:model_CGLS}, respectively. A complex-valued clipping function ${\mathcal{Q}}_{\rm{clip}}$ is defined as ${\mathcal{Q}}_{\rm{clip}}\equiv Q_{\rm{clip}}({\rm{Re}}\{z_m + n_m\})+jQ_{\rm{clip}}({\rm {Im}}\{z_m + n_m\})$, $m=1, ..., M$, i.e., the real and imaginary parts are separately clipped symbol by symbol as follows.
\BE\label{Eqn:clip_func}
Q_{\rm{clip}}(x) {=}     \begin{cases}
        \lambda,       & {\mr {if}}\;{x \ge \lambda} \\%[1.5mm]
        x,     & {\mr {if}}\;{- \lambda < x < \lambda} \\%[1.5mm]
        - \lambda,     & {\mr {if}}\;{x \le  - \lambda}
    \end{cases}.
\EE  
with $\lambda$ being the clipping threshold. Since the clipped signal is detected independently symbol-by-symbol, we ignore the subscript $m$ to simplify the discussion.
According to Fig.~\ref{Fig:Var_GOAMP}, noting that $\bf{\bb{z}}_t-z-y$ is a Markov chain, we have
\BE\label{Eqn:pz_1}
P(z|\bf{\bb{z}}_t,y) = \frac{P(\bf{\bb{z}}_t|z)P(y|z)}{\int P(\bf{\bb{z}}_t|z)P(y|z)dz},
\EE
As a result, given $\lambda$, the \emph{a posteriori} mean of $z$ is
\BS\label{Eqn:clip_z}
\BE\label{Eqn:z_post}
\hat{z}=\hat{\psi}_t(\bf{\bb{z}}_t)={\mr E}\{z|\bf{\bb{z}}_t,y\}=\int zP(z|\bf{\bb{z}}_t,y)dz,
\EE
and the \emph{a posteriori} variance of $z$ is 
\BE\label{Eqn:vz_post}
\hat{v}_t^z={\mr E}\{z^2|\bf{\bb{z}}_t,y\}-\hat{z}^2.
\EE
\ES
Based on the \emph{a posteriori} estimation $(\hat{z}, \hat{v}_t^z)$, \eqref{Eqn:SE_IIGOAMP},  and Theorem~\ref{them:capacity}, we can derive the achievable rate of GOAMP/GVAMP with de-clipping.

\section{Achievable Rates of MRC in linearized model}\label{APP:MRC_Rate}%}
In the existing works\cite{ShansuoWCL2019,Tong2010_clip}, to simplify the analysis of clipping, a linearized approximate clipping model is considered, i.e, 
\BE\label{Eqn:linear_clip}
\begin{cases}
\bf{y}=\alpha {\bf{z}} + \bf{d},\\
\bf{z}= \bf{A}\bf{x}+\bf{n},
\end{cases}
\EE
where $\alpha$ is a constant scalar computed as $\alpha = \frac{{\mr{E}}[\bf{z}^H{\bf{y}}]}{{\mr E}[||\bf{z}||^2]}$ and $\bf{d}=\bf{y}-\alpha\bf{z}$ is the clipping distortion. Meanwhile, the MRC receiver is commonly used in wireless communications~\cite{wck2015ADC}. The achievable rate of MRC receiver with Gaussian signaling is given by \cite{wck2015ADC}
\BE\nonumber
 \!\!\!\!R_{\text{MRC}}\!\!=\!\!\sum_{m=1}^{M}{\mr{log}}_2\big(1+\tfrac{\alpha^2 \mr{E}\{||\bf{h}_m||^4\}}{\alpha^2\sum_{i\neq m}\mr{E}\{|\bf{h}_m^H \bf{h}_i|^2\}+(\alpha^2\sigma^2+||\bf{d}||^2){\mr{E}\{||\bf{h}_m||^2\}}}\big). 
\EE 

\section{Optimized irregular LDPC codes with GOAMP/GVAMP for clipping}\label{APP:Opt_code}%}
\begin{table}[!h] \scriptsize \caption{Optimized irregular LDPC codes with GOAMP/GVAMP for clipping.}\label{Opt_degree1}
	\centering\setlength{\tabcolsep}{0.2mm}{
		\begin{tabular}{|c||c|c|}
			\hline
			\multicolumn{1}{|c||}{\multirow {2}{*}{\tabincell{c}{ System\\  parameters}}}&\multicolumn{1}{c|}{$N=M$} &\multicolumn{1}{c|}{\tabincell{c}{target $R_{\rm{sum}}$}}\\
			\cline{2-3}
			\multicolumn{1}{|c||}{} & \multicolumn{1}{c|}{500} & \multicolumn{1}{c|}{500} \\
			\hline
		\multicolumn{1}{|c||}{\multirow {2}{*}{Scenarios}}& \multicolumn{2}    {c|}{\multirow {1}{*}{clipping ($\lambda=1$)}} \\
			\cline{2-3}
			\multicolumn{1}{|c||}{} & \multicolumn{1}{c|}{$\kappa=10$} & \multicolumn{1}{c|}{$\kappa=50$}\\
			\hline 
                ${\rm{Code~length}}$ & \multicolumn{2}{c|}{$10^5$}\\
                \hline
			 $R_{\mr{LDPC}}$ & 0.5 & 0.5\\
			\hline
			$\mu(X)$&{\tabincell{c}{ ${\it{\mu}}_{\text{6}}=1$ }}&\multicolumn{1}{c|}{ ${\it{\mu}}_{\text{8}}=1$}\\
			\hline
			\multicolumn{1}{|c||}{\multirow {6}{*}{$\gamma(X)$}} & $\gamma_{2}=0.4604$ & $\gamma_{2}=0.4619$\\
			\multicolumn{1}{|c||}{} & $\gamma_{3}=0.2464$    & $\gamma_{14}=0.0196$      \\
			\multicolumn{1}{|c||}{} & $\gamma_{13}=0.1743$    & $\gamma_{15}=0.2559$      \\  
			\multicolumn{1}{|c||}{} & $\gamma_{14}=0.1189$    & $\gamma_{70}=0.0956$     \\    
			\multicolumn{1}{|c||}{} &     & $\gamma_{80}=0.0760$   \\  
			\multicolumn{1}{|c||}{} &     & $\gamma_{500}=0.0910$ \\
	    \hline
		$(snr)^{\it{\ast}}_{\text{dB}}$ & \multicolumn{1}{c|}{2.25} & \multicolumn{1}{c|}{3.7} \\
			\hline
	    ${\text{(limit)}_{\text{dB}}}$ & \multicolumn{1}{c|}{2.14} & \multicolumn{1}{c|}{3.68} \\
			\hline
	\end{tabular}}\vspace{-0.2cm}
\end{table}
Similar as \cite[Section A]{LeiTIT2021},  a kind of practical irregular LDPC code $(\bf{\gamma}(X)=\sum_{i=2}^{d_{v,\mr{max}}}\gamma_i X^{i-1}$, $\bf{\mu}(X))=\sum_{i=2}^{d_{c,\mr{max}}}\mu_i X^{i-1})$ is optimized for GOAMP/GVAMP with clipping ($\lambda=1$), where $d_{v,\mr{max}}$ and $d_{c,\mr{max}}$ are the corresponding maximum degrees of VN and CN. In Table \ref{Opt_degree1},  the optimized irregular LDPC codes are given with target rate $R_{\mr{sum}}=NR_{\mr{LDPC}}{\mr{log}}_{2}|\mathcal{S}_{\mr{QPSK}}|=500$, where $R_{\mr{LDPC}}=0.5$ and $|\mathcal{S}_{\mr{QPSK}}|=4$. Note that the decoding thresholds of the optimized LDPC codes are within $0.11$ dB away from the theoretical limits for the maximum achievable rate of GOAMP/GVAMP.

To verify the advantages of the optimized LDPC codes for GOAMP/GVAMP, we employ P2P well-designed $R_{\rm{LDPC}}=0.5$ with the degree distributions $\lambda(X)=0.24426x+0.25907x^2+0.01054x^3+0.05510x^4+0.01455^7+0.01275x^9+0.40373x^{11}$ and $\mu(X)=0.25475x^6+0.73438x^7+0.01087x^8$ \cite{Richardson2001} as the baseline method, whose the decoding threshold is $0.18$~dB away from the P2P-AWGN capacity.

\end{appendices}

\end{document}